\documentclass[11pt,onecolumn]{IEEEtran}
\pdfoutput=1
\usepackage{cite}
\usepackage{amsmath,amssymb,amsfonts}
\usepackage{scalerel,stackengine}
\stackMath
\newcommand\reallywidehat[1]{%
\savestack{\tmpbox}{\stretchto{%
  \scaleto{%
    \scalerel*[\widthof{\ensuremath{#1}}]{\kern-.6pt\bigwedge\kern-.6pt}%
    {\rule[-\textheight/2]{1ex}{\textheight}}
  }{\textheight}%
}{0.5ex}}%
\stackon[1pt]{#1}{\tmpbox}%
}
\usepackage{algorithm}
\usepackage[english]{babel}
\usepackage[utf8]{inputenc}
\usepackage[noend]{algpseudocode}
\usepackage{graphicx}
\usepackage{textcomp}
\usepackage{setspace}
\usepackage [english]{babel}
\usepackage [autostyle, english = american]{csquotes}
\MakeOuterQuote{"}
\def\BibTeX{{\rm B\kern-.05em{\sc i\kern-.025em b}\kern-.08em
    T\kern-.1667em\lower.7ex\hbox{E}\kern-.125emX}}
    
\begin{document}
\title{Learning-based physical layer communications for multi-agent collaboration}
\author{\IEEEauthorblockN{Arsham Mostaani$\, ^1$, Osvaldo Simeone$\, ^2$, Symeon Chatzinotas$\, ^1$ and Bjorn Ottersten$\, ^1$}

\IEEEauthorblockA{\normalsize{{$^1$ SIGCOM laboratory, SnT Interdisciplinary Centre, \\University of Luxembourg, Luxembourg} \\
{$^2$ Centre for Telecom. Research, Department of Informatics,\\ King's College London, London, UK}\\
\{arsham.mostaani,symeon.chatzinotas,bjorn.ottersten\}@uni.lu , \{osvaldo.simeone\}@kcl.ac.uk}}

}
\maketitle

\vspace{-2.0cm}
\begin{abstract}
Consider a collaborative task carried out by two autonomous agents that can communicate over a noisy channel. Each agent is only aware of its own state, while the accomplishment of the task depends on the value of the joint state of both agents. As an example, both agents must simultaneously reach a certain location of the environment, while only being aware of their own positions. Assuming the presence of feedback in the form of a common reward to the agents, a conventional approach would apply separately: (\emph{i}) an off-the-shelf coding and decoding scheme in order to enhance the reliability of the communication of the state of one agent to the other;  and (\emph{ii}) a standard multi-agent reinforcement learning strategy to learn how to act in the resulting environment. In this work, it is argued that the performance of the collaborative task can be improved if the agents learn how to jointly communicate and act. In particular, numerical results for a baseline grid world example demonstrate that the jointly learned policy carries out compression and unequal error protection by leveraging information about the action policy.
\end{abstract}
\section{Introduction}
\label{sec:introduction}

Consider the rendezvous problem illustrated in Fig. 1 and Fig. 2. Two agents, e.g., members of a SWAT team, need to arrive at the goal point in a grid world at precisely the same time, while starting from arbitrary positions. Each agent only knows its own position but is allowed to communicate with the other agent over a noisy channel. This set-up is an example of cooperative multiple agent problems in which each agent has partial information about the environment \cite{pynadath2002communicative,weiss1999multiagent}. In this scenario, communication and coordination are essential in order to achieve the common goal \cite{tan1993multi,busoniu2008comprehensive,lauer2000distributedQ}, and it is not optimal to design the communication and control strategies separately \cite{lauer2000distributedQ, SahaiDemystifying}.

\vspace{-0.0cm}
 \begin{figure}[b]\label{multi-agent}
  \centering

      \includegraphics[width=0.7\textwidth]{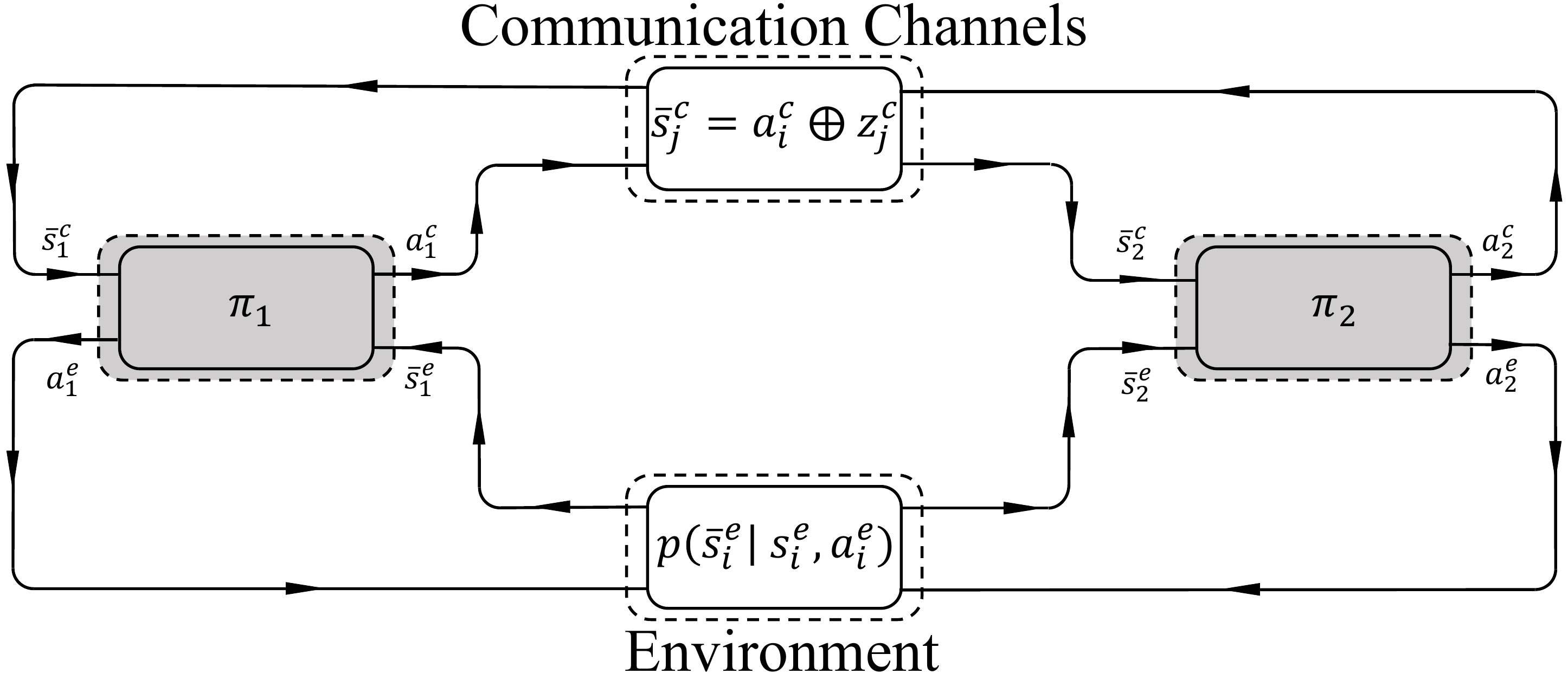}
  \caption{An illustration of the cooperative multi-agent system.}
\end{figure}

 \begin{figure*}[t!]\label{rendez-vous}
  \centering

      \includegraphics[width=0.75\textwidth]{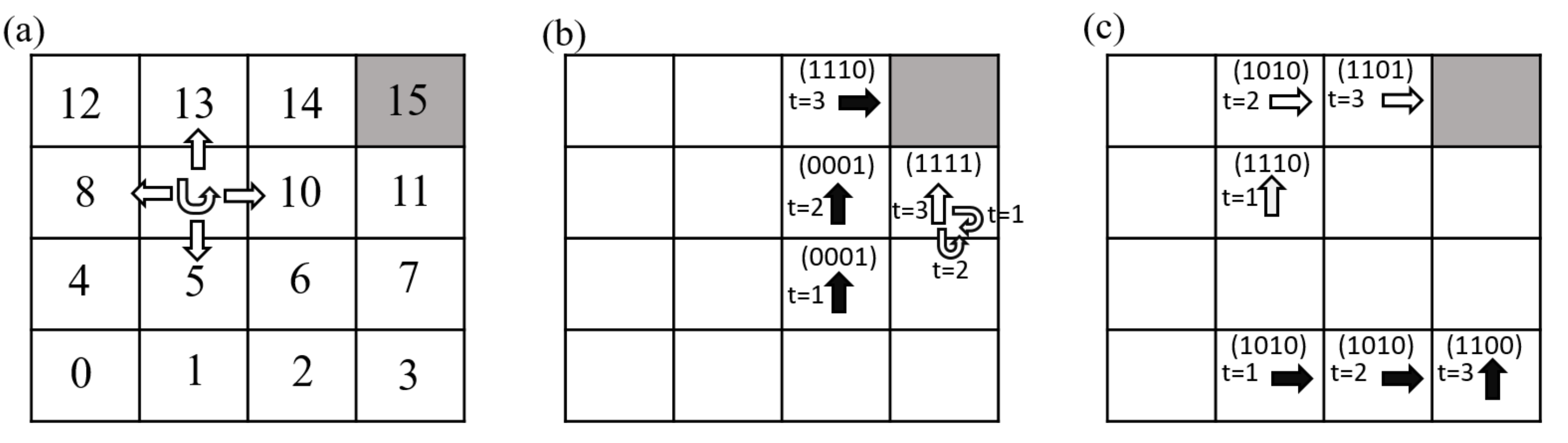}
  \caption{The rendezvous problem: (a) illustration of the environment state-space, $ \mathcal{S}^e$, i.e., the location on the grid, of the environment action space $\mathcal{A}^e$, denoted by arrows, and of the goal state, marked with gray background; (b) demonstration of a sampled episode, where arrows show the environment actions taken by the agents (empty arrows: actions of agent 1, solid arrows: actions of agent 2) and the $B=4$ bits represent the message sent by each agent. A larger reward $R_2> R_1$ is given to both agents when they enter the goal point at the same time, as in the example; (c) in contrast, $R_1$ is the reward accrued by agents when only one agent enters the goal position.}
\end{figure*}


Assuming the presence of a delayed and sparse common feedback signal that encodes the team reward, cooperative multi-agent problems can be formulated in the framework of multi-agent reinforcement learning. As attested by the references \cite{pynadath2002communicative,weiss1999multiagent,fischer2004hierarchical} mentioned above, as well by \cite{sukhbaatar2016learning,FoersterLearning}, this is a well-studied and active field of research. To overview some more recent contributions, paper \cite{DoyaEmergence} presents simulation results for a distributed tabular Q-learning scheme with instantaneous communication. Deep learning approximation methods are applied in \cite{FoersterStabilising} for Q-learning and in \cite{LoweActorCritic} for actor-critic methods. In \cite{FoersterCounter}, a method is proposed that keeps a centralized critic in the form of a Q-function during the learning phase and uses a counter-factual approach to carry out credit assignment for the policy gradients. 
 
The works mentioned above assume a noiseless communication channel between agents or use noise as a form of regularization \cite{FoersterLearning}. In contrast, in this paper, we consider the problems of simultaneously learning how to communicate on a noisy channel and how to act, creating a bridge between the emerging literature on machine learning for communications \cite{simeone2018very} and multi-agent reinforcement learning. 

Our specific contributions are as follows. First, we formulate distributed Q-learning
algorithms that learn simultaneously what to communicate on a noisy channel and which actions to take in the environment in the presence of communication delays. Second, for the rendezvous problem illustrated in Fig. 2, we provide a numerical performance comparison between the proposed multi-agent reinforcement learning scheme and a conventional method. The proposed scheme jointly learns how to act and communicate, where the conventional method applies separately an off-the-shelf channel coding scheme for communication and multi-agent reinforcement learning to adopt the action policies. Unlike the conventional method, the jointly optimized policy is seen to be able to learn a communication scheme that carries out data compression and unequal error protection as a function of the action  policy.

\section{Problem Set-up}\label{problem set-up}
As illustrated in Fig. 1 and Fig. 2, we consider a cooperative multi-agent system comprising of two agents that communicate over noisy channels. The system operates in discrete time, with agents taking actions and communicating in each time step $t=1,2,...$. While the approach can be applied more generally, in order to fix the ideas, we focus here on the rendezvous problem illustrated in Fig. 2. The two agents operate on an $n \times n$ grid world and aim at arriving at the same time at the goal point on the grid. The position of each agent $i \in \{1,2\}$ on the grid determines its environment state $s^e_{i} \in \mathcal{S}^e = [n] \times [n]$, where $[n]=\{1,2,...,n\}$. Each agent $i$th environment state $s^e_{i} \in \mathcal{S}^e$ can also be written as the pair $s^e_{i}=\langle s^e_{i,x} ,s^e_{i,y} \rangle$, with $s^e_{i,x} \, , s^e_{i,y} \in [n]$ being respectively the horizontal and vertical coordinates. Each episode terminates as soon as an agent or both visit the goal point which is denoted as $\mathcal{S}^e_T=\{ s^e_T \}$. At time $t=1$, the initial position $\mathrm{s}_{i,t=1}^e$, is randomly and uniformly selected amongst the non-goal states. Note that, throughout, we use Roman font to indicate random variables and the corresponding standard font for their realizations.
 
At any time step $t=1,2,...$ each agent $i$ has information about its position, or environment state, $s^e_{i,t}$ and about the signal $s^c_{i,t}$ received from the other agent $j \neq{i}$ at the previous time step $t-1$.  Based on this information, agent $i$ selects its environment action $a_i^e=\langle a^e_{i,x} ,a^e_{i,y} \rangle$ from the set $\mathcal{A}^e = \{\langle 1,0 \rangle,\langle-1,0 \rangle,\langle 0 ,0 \rangle,\langle 0 , 1 \rangle,\langle 0 , -1\rangle\}$, where $a^e_{i,x}$ and $a^e_{i,y}$ represent the horizontal and vertical move of agent $i$ on the grid. 
Furthermore, it chooses the communication message to send to the other agent by selecting a communication action $a^c_i \in \mathcal{A}^c = \{0,1\}^B$ of $B$ bits. 

The environment state transition probability 
for agent $i$ can be described by the equation
$
\mathrm{s}^e_{i,t+1}= \mathrm{s}^e_{i,t}+\mathrm{a}^e_{i,t} \,
$,
with the caveat that, if an agent on an edge of the grid world selects an action that transfers it out, the environment keeps the agent at its current location. Agents communicate over interference-free channels using binary signaling, and the channels between the two agents are independent Binary Symmetric Channels (BSCs), such that the received signal is given as
\begin{equation} \label{channel model - specific}
\mathrm{s}^c_{j , t+1} = {a}^c_{i ,t} \oplus \mathrm{z}^c_{j , t},  
\end{equation}
where the XOR operation $\oplus$ is applied element-wise, and $\mathrm{z}^c_{j,t}$ has independent identically distributed (i.i.d.) Bernoulli entries with bit flipping probability $q \leq 0.5$.


Each agent $i$ follows a policy $\pi_i$ that maps the observations $s_i=\langle s^e_i, s^c_i \rangle$ of the agent into its actions $a_i=\langle a^e_i, a^c_i \rangle$. The policy is generally stochastic, and we write it as the conditional probability $\pi_i(a_i|s_i)$ of taking action $a_i$ while in state $s_i$.
We assume the policy $\pi_i$ to be factorized as
\begin{equation}\label{Policy decomposition}
    \pi_i(a_i|s_i) = \pi_i^e(a_i^e|s_i^e, s_i^c) \, \pi_i^c(a_i^c|s_i^e),  
\end{equation}
into a component selecting the environment action $a^e_i$ based on the overall state $s_i$ and one selecting the transmitted signal $a^c_i$ based on the current position $s^e_i$. The overall joint policy $\pi$ is given by the product $\pi = \pi_1 \times \pi_2$. It is noted that the assumed memoryless stationary policies are sub-optimal under partial individual observability of environment state \cite{pynadath2002communicative}.

At each time $t$, given states $\langle s_1,s_2 \rangle$ and actions $\langle a_1,a_2 \rangle$, both agents receive a single team reward \begin{equation}\label{example environment noise distribution}
   \mathrm{r}_t =
   \begin{cases}
    R_1, & \text{if  ${s}^{e}_{i} \neq {s}^{e}_{j} \in \mathcal{S}_T^e$}\\
    R_2, & \text{if  ${s}^{e}_{i} = {s}^{e}_{j} \in \mathcal{S}_T^e$},\\
    0, & \text{otherwise},\\
   \end{cases}
\end{equation}
where $R_1 < R_2$. Accordingly, when only one agent arrives at the target point $s^e_T$, a smaller reward $R_1$ is obtained at the end of the episode, while the larger reward $R_2$ is attained when both agents visit the goal point at the same time. The goal of the multi-agent system is to find a joint policy $\pi $ that maximizes the expected return. For given initial states, $({s}_{1,t=1}, s_{2,t=1})$, this amounts to solving the problem
\begin{equation} \label{main problem}
\begin{aligned}
& \underset{\pi}{\text{maximize}}
\quad \quad {\mathbb{E}_{\pi}[\mathrm{G}_{t}| \mathrm{s}_{1,t=1}={s}_{1,t=1}, \mathrm{s}_{2,t=1}=s_{2,t=1}]}, \quad 
\end{aligned}
\end{equation}  
where
\begin{equation}\label{Longterm return}
  \mathrm{G}_{t}= \sum_{t=1}^{\infty} \gamma^t \mathrm{r}_t \quad   
\end{equation}
is the long-term discounted return, with $\gamma \in (0,1]$ being the reward discount factor. The expected return in (\ref{main problem}) is calculated with respect to the probability of the trace of states, actions, and rewards induced by the policy $\pi$ \cite{Suttonintroduction}.

\section{Learned Communication}\label{Learned Communication}

In this section we consider a strategy that jointly learns the communication and the environment action policies of both agents, by tackling problem (\ref{main problem}). To this end, we apply the policy decomposition (\ref{Policy decomposition}) and use the distributed Q-learning algorithm \cite{lauer2000distributedQ}. Accordingly, given the received communication signal $s^c_i$ and the local environment state $s^e_i$, each agent $i$ selects its environment actions $a^e_i$ by following a policy $\pi^e_i$ based on a state-action value function $Q^e_i(s^e_i,s^c_i,a^e_i)$; and it chooses its communication action $a^c_i$ by following a second policy $\pi^c_i$, based on a state-action function $Q^c_i(s^e_i,a^c_i)$. We recall that a state-action function $Q(s,a)$ provides an estimate of the expected return (\ref{Longterm return}) when starting from the state $s$ and taking action $a$.

In order to control the trade-off between exploitation and exploration, we adopt the Upper Confidence Bound (UCB) method \cite{Suttonintroduction}. UCB selects the communication action $\mathrm{a}^c_{i,t}$ as
\begin{equation}
     {a}^c_{i,t} = \underset{a^c_i}{\text{argmax}}  \,\, Q^c_i(s^e_{i,t},a^c_{i}) + c \sqrt{\frac{\mathrm{ln}(T_t)}{N_{i}^c(s_{i,t}^e,a^c_{i})}},
\end{equation}
where $c>0$ is a constant;  $T_t$ is the total number of time steps in the episodes considered up to the current time $t$ in a given training epoch; and table $N_{i}^c(s_{i,t}^e,a^c_{i})$ counts the total number of times that the state $s^e_{i,t}$ has been visited and the action $a^c_i$ selected among the previous $T_t$ steps. When $c$ is large enough, UCB encourages the exploration of the state-action tuples that have been experienced fewer times. A similar rule is applied for the environment actions $a^e_i$.

The update of the Q-tables follows the off-policy Q-learning algorithm, i.e., \vspace{-0.5cm}

\begin{equation}\label{Qe update}
     Q^e_{i}(s_{i,t}^e,s_{i,t}^c,a_{i,t}^e)     \leftarrow
     \! \,(1-\alpha)Q^e_{i}(s_{i,t}^e,s_{i,t}^c,a_{i,t}^e) + \alpha \gamma \Big(\! \mathrm{r_t}+ \underset{{a}_{i}^e}{\text{max}}\, Q^e_{i}(s_{i,t+1}^e,s_{i,t+1}^c,a_{i}^e)\!\Big)
\end{equation}

\vspace{-0.7cm}
\begin{equation}\label{Qc update}
    Q^c_{i}(s_{i,t}^e,a_{i,t}^c) \leftarrow
     \! \,(1-\alpha)Q^c_{i}(s_{i,t}^e,a_{i,t}^c) + \alpha \gamma \Big(\mathrm{r_t}+ \underset{{a}_{i}^c}{\text{max}}\, Q^c_{i}(s_{i,t+1}^e,a_{i}^c)\Big),
\end{equation}
where $\alpha>0$ is a learning rate parameter. The full algorithm is detailed in Algorithm 1.

As a baseline, we also consider a conventional communication scheme, whereby each agent $i$ sends its environment state $s^e_i$ to the other agent by using a channel code for the given noisy channel. Agent $j$ obtains an estimate $\hat{s}^e_i$ of the environment state of $i$ by using a channel decoder. This estimate is used as if it were the correct position of the other agent to define the environment state-action value function $Q^e_j (s^e_j,\hat{s}^e_i,a^e_j)$. This table is updated using Q-learning and the UCB policy in a manner similar to Algorithm 1.


\begin{algorithm}\label{DCQ SCA Delayed UCB}
\caption{Learned Communication}\label{alg:euclid}
\begin{algorithmic}[1]
\State \small \textbf{Input:} $\gamma$ (discount factor), $\alpha$ (learning rate), and $c$ (UCB exploration constant)

 \State \textbf{Initialize} all zero Q-tables $Q^e_{i}(s_{i}^e,s^c_i,a^e_{i})$ and  $Q^c_{i}(s^c_i,a^c_{i})$, and tables $N_{i}^e(s_{i}^e,s^c_i,a^e_{i})$ and $N_{i}^c(s^c_i,a^c_{i})$, for $i=1,2$ 
  \For{each episode $m=1:M$}
 \State Randomly initialize $\langle s^e_{1,t=1},s^e_{2,t=1}\rangle$ and $\langle s^c_{1,t=1},s^c_{2,t=1}\rangle$
 \State set $t_m=1$
\While{$\langle s^e_{1,t} , s^e_{2,t}\rangle  \notin \mathcal{S}^e_T$}
        \State Select ${a}_{i,t}^c = a^c_i \in \mathcal{A}_i^c$, that maximizes 
        
        $\quad Q^c_i(s^c_{i,t},a^c_{i}) + c \sqrt{\frac{\mathrm{ln}(\sum_{k=1}^m t_k)}{N_{i}^c(s^c_{i,t},a^c_{i})}}$, {for} $i=1,2$
\vspace{1mm}
        \State Update $N_{i}^c(s^c_{i,t},a^c_{i,t}) \leftarrow N_{i}^c(s_{i,t}^c,a^c_{i,t}) + 1$

\vspace{1mm}          
            \State Select $a^e_{i,t} = a^e_i \in \mathcal{A}_i^e$ that maximizes

            $\quad Q^e_i(s^e_{i,t},s^c_{i,t},a^e_{i})+ c \sqrt{\frac{\mathrm{ln}(\sum_{k=1}^m t_k)}{N_{i}^e(s_{i,t}^e,s^c_{i,t},a^c_{i})}}$, for $i=1,2$ 
\vspace{1mm}
            \State Update $N_{i}^e(s_{i,t}^e,s^c_{i,t},a^e_{i,t}) \leftarrow N_{i}^e(s_{i,t}^e,s^c_{i,t},a^e_{i,t}) + 1$
\vspace{1mm}
            \State Obtain message $s^c_{i,t+1}$, {for} $i=1,2$
\vspace{1mm}
            \State Obtain ${r}_t$ and move to  $s^e_{i,t+1}$, {for} $i=1,2$
    \vspace{1mm}
    
    \For{$i=1,2$}
        \State Update $Q^e_{i}(s_{i,t}^e,s_{i,t}^c,a_{i,t}^e)$ following (\ref{Qe update}) 
        
        
        \State Update $Q^c_{i}(s_{i,t}^c,a_{i,t}^c)$ following (\ref{Qc update}) 
        
    \EndFor 
    \textbf{end}
            
\State $t_m=t_m+1$

\EndWhile\label{euclidendwhile}
\State \textbf{end}
\EndFor
\State Compute $\sum^{t_m-1}_{t=1} \gamma^t \mathrm{r}_t$ for the $m$th episode
\State \textbf{end}
\vspace{1mm}
\State
\State \textbf{Output:} $ \pi^e_i(a^e_i|s^e_i,s^c_i)=\delta\Big( a^e_i -
 \underset{{a}_{i}^e \in \mathcal{A}^e_i}{\text{argmax}} \, Q^{e}_i(s^e_i,{s}^c_i,a^e_i)\Big)
  $ and
  
  $\; \; \; \pi^c_i(a^c_i|s^e_i)= \delta\Big( a^c_i -
 \underset{{a}_{i}^c \in \mathcal{A}^c_i}{\text{argmax}} \, Q^{c}_i(s^c_i,a^c_i)
  \Big)$  for $i=1,2$
%
\end{algorithmic}
\end{algorithm}

\vspace{-0.4cm}\section{Results and Discussions}\label{results Section}
 In this section, we provide numerical results for the rendezvous problem described in Sec. \ref{problem set-up}. As in Fig. 2, the grid world is of size $4 \times 4$, i.e. $n=4$, and it contains one goal point at the right-top position. Environment states are numbered row-wise starting from the left-bottom as shown in Fig. 2(a). All the algorithms are run for $50$ independent epochs. For each agent $i$ the initial state $ s^e_{i,t=1} \notin \mathcal{S}^e_T$ in each episode is drawn uniformly from all non-terminal states.


We compare the conventional communication and the learned communication schemes reviewed in the previous section. Conventional communication transmits the position of an agent on the grid as the 4-bit binary version of the indices in Fig. 2(a) after encoding via a binary cyclic ($B$,4) code, where the received message is decoded by syndrome decoding.

The performance of each scheme is evaluated in terms of the discounted return in (\ref{Longterm return}), averaged over all epochs and smoothed using a moving average filter of memory equal to 4,000 episodes. The rewards in (\ref{example environment noise distribution}) are selected as $R_1=1$ and $R_2=3$, while the discount factor is $\gamma = 0.9$. A constant learning rate $\alpha=0.15$ is applied, and the exploration rate $c$ of the UCB policy is selected from the set $\{ 0.3, 0.4, 0.5 \}$ such that it maximizes the average return at the end of the episodes in an epoch.

We first investigate the impact of the channel noise by considering different values of the bit flip probability $q$. In Fig. 3 it is observed that conventional communication performs well at the low bit flipping rate of $q=0.05$, but at higher rates of $q$ learned communication outperforms conventional communication after a sufficiently large number of episodes. Importantly, for $q= 0.2$, the performance of conventional communication degrades through episodes due to the accumulation of noise in the observations, while learned communication is seen to be robust against channel noise.


 \vspace{0cm}
 \begin{figure}[t]\label{Compare learning and convent - no delay - q03q06q09}
  \centering
  









      \includegraphics[width=0.5\textwidth]{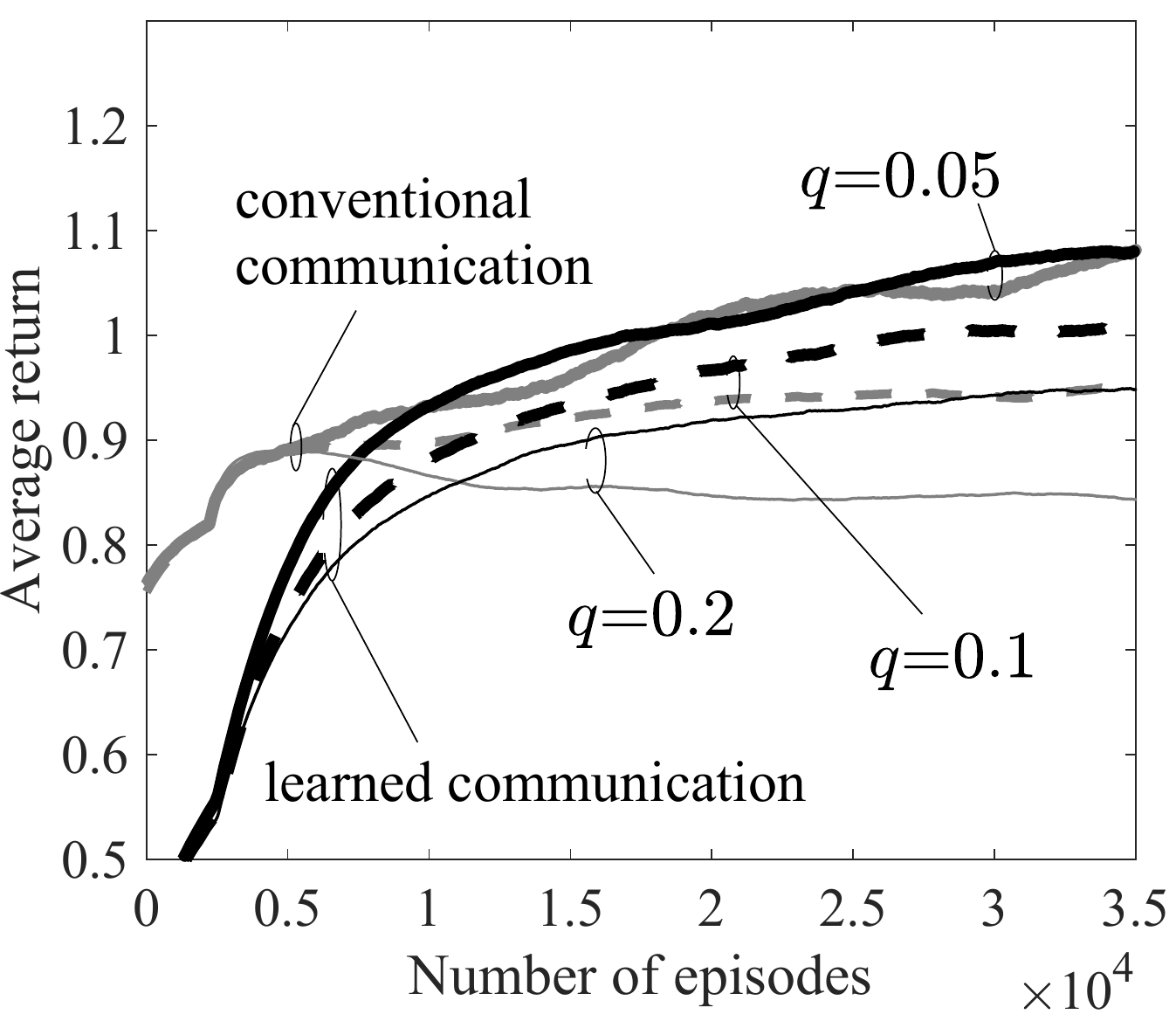}
  \vspace{0.0cm}      
  \caption{Average return for conventional communication and learned communication when $B=7$.}
  \vspace{-0.5cm}
\end{figure}

We now discuss the reasons that underlie the performance advantages of learned communication. We start by analyzing the capability of learned communication to compress the environment state information before transmission. To obtain quantitative insights, we measure the mutual information $I({\mathrm{s}^e_i;\mathrm{a}^c_i})$ between the environment state $\mathrm{s}^e_i$ and the communication action $\mathrm{a}^c_i$ of an agent $i$ as obtained under the policy learned after 20,000 episodes for $q = 0,0.05,0.1,0.15,0.2$. Fig. 4 plots the mutual information as a function of the bit flipping probability $q$ for learned communication. For conventional communication scheme the communication message $a^c_i$ is a deterministic function of the state $s^e_i$ and hence we have $I(\mathrm{s}^e_i;\mathrm{a}^e_i) = H(\mathrm{s^e_i})$, which is independent of $q$ and $B$. In the absence of channel noise, i.e., $q=0$, learned communication compresses by almost 30\% the information about the environment state distribution $\mathrm{s}^e_i$ when $B=6$. This reduction becomes even more pronounced as the channel noise increases or when agents have a tighter bit-budget.

 \begin{figure}[t]\label{Compare mutual information}
  \centering

      \includegraphics[width=0.5\textwidth]{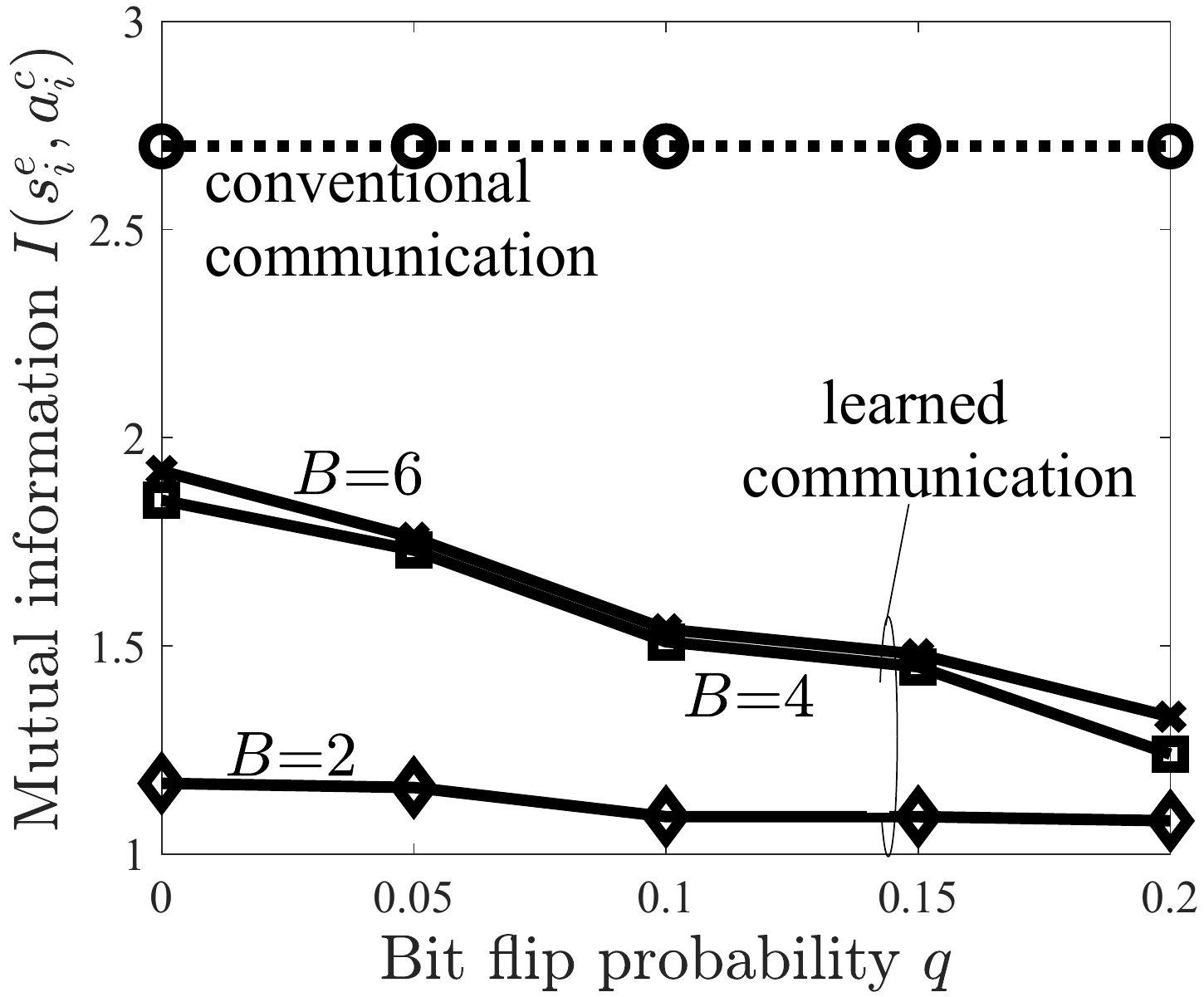}
  \caption{Mutual information between an agent's environment state $\mathrm{s}^e_i$ and the communication action $\mathrm{a}^c_i$ versus the bit flip probability $q$ for conventional communication and learned communication with delay after 20,000 episodes ($B=2,4,6$,\, $q=0.05,0.10,0.15,0.20$).}
  \vspace{-0.0cm}
\end{figure}

We proceed by investigating how compression is carried out by jointly optimizing the agent's environment action  and communication action policies. We will also see that learned communication carries out a form of unequal error protection. To this end, Fig. 5 illustrates a sample of the learned action and communication policies $\pi^e_i$ and $\pi^c_i$ for agent $i=1$ when $q=0.05$ and $B=4$ after 30,000 episodes of training in the presence of communication delays. In this figure, arrows show the dominant environment action(s) $a^e_{i}$ selected at each location; the bit sequences represent the communication action $a^c_i$ selected at each location; and the colour of each square shows how likely it is for the position to be visited by agent $i$.

 \vspace{0cm}
 \begin{figure}[t!]\label{Gridworld-example-environment state distribution}
  \centering

      \includegraphics[width=0.5\textwidth]{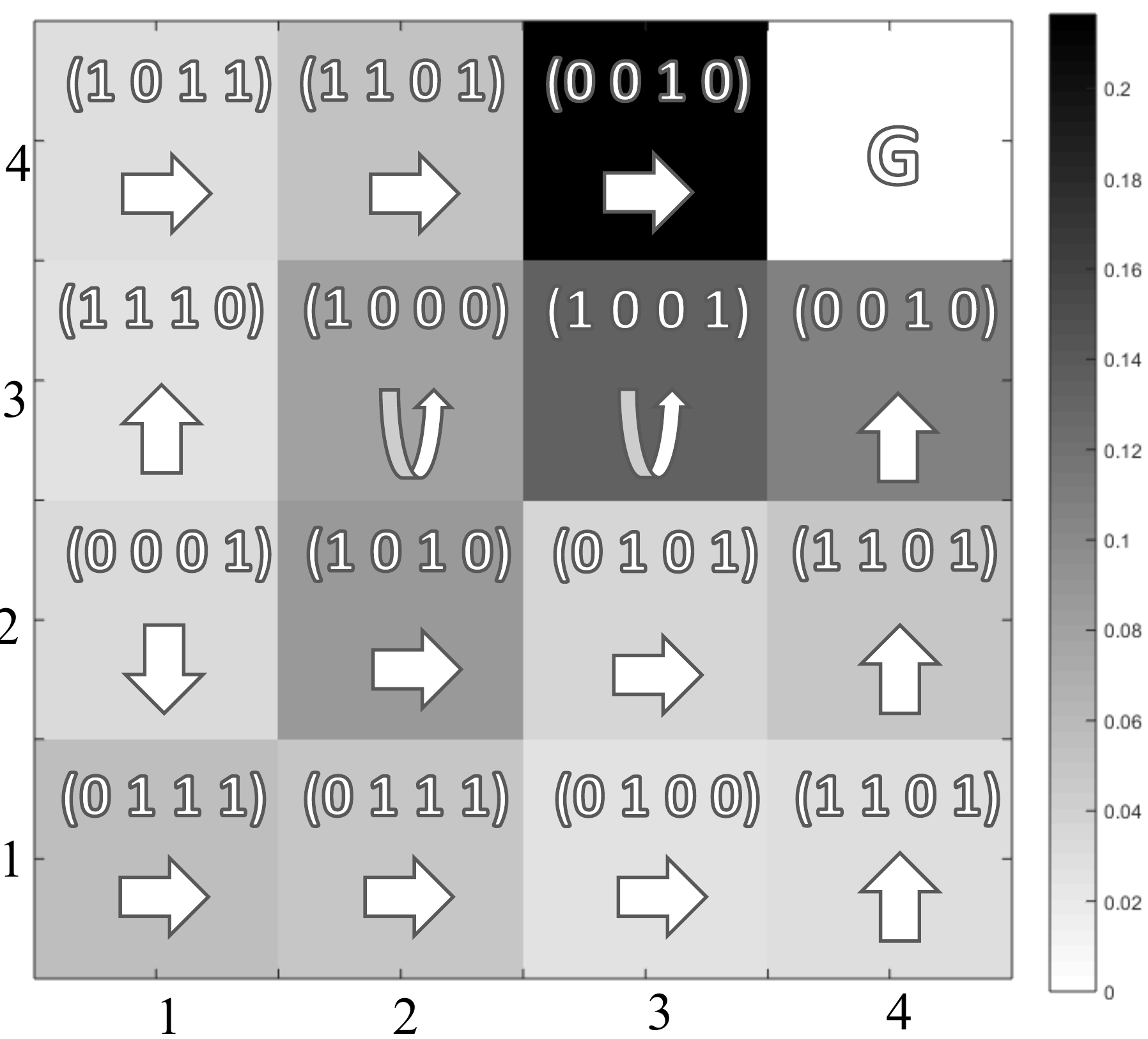}
  \vspace{-0.0cm}  
  \caption{ Illustration of a learned communication action policy when there is no communication delay ($B=4$, $q=0.05$). Locations with brighter colors are more likely to be visited. Arrows show the dominant action selected at any location. Bit strings show the message sent at a certain location.}
  \vspace{0.1cm}
\end{figure}

We can observe that compression is achieved by assigning same message to different locations. In this regard, it is interesting to note the interplay with the learned action policy: groups of states are clustered together if states have similar distance from the goal point, such as $\{\langle 4,3 \rangle , \langle 3,4\rangle \}$ and $\{\langle 4,2 \rangle , \langle 2,4\rangle \}$; or if they are very far from the goal point such as $\{\langle 1,2 \rangle , \langle 1,3 \rangle \}$.
 Furthermore, it is seen that the Hamming distance of the selected messages depends on how critical it is to distinguish between the corresponding states. This is because it is important for an agent to realize whether the other agent is close to the terminal point.
 
 \vspace{-0.1cm}\section{Conclusions}\label{conclusion section}
In this paper we have studied the problem of decentralized control of agents that communicate over a noisy channel. The results demonstrate that jointly learning communication and action policies can significantly outperform methods based on standard channel coding schemes and on the separation between the communication and control policies. We observed this performance gain for delayed and noisy inter-agent communication and we discussed that the underlying reason for the improvement in performance is the learned ability of the agents to carry out data compression and unequal error protection as a function of the action policies.

\section{Acknowledgements}
The work of Arsham Mostaani, Symeon Chatzinotas and Bjorn Ottersten is supported by European Research Council (ERC) advanced grant 2022 (Grant agreement ID: 742648).
Arsham Mostaani and Osvaldo Simeone have received funding from the 
European Research Council (ERC) under the European Union's Horizon 2020 
Research and Innovation Program (Grant Agreement No. 725731).

\vspace{-0.1cm}

\end{document}